\documentclass[12pt]{article}
\usepackage{epsfig,cite}
\def\gsim{\:\raisebox{-0.8ex}{$\stackrel{\textstyle>}{\sim}$}\:}
\def\lsim{\:\raisebox{-0.8ex}{$\stackrel{\textstyle<}{\sim}$}\:}

\newcommand{\ben}{\begin{eqnarray}\displaystyle}
\newcommand{\een}{\end{eqnarray}}
\newcommand{\refb}[1]{(\ref{#1})}

\newcommand{\sectiono}[1]{\section{#1}\setcounter{equation}{0}}

\def\beq{\begin{equation}}
\def\eeq{\end{equation}}
\def\barr{\begin{array}}
\def\earr{\end{array}}
\def\dis{\displaystyle}

\begin{document}

\thispagestyle{empty}

\begin{flushright}
{\large \tt hep-th/0305104}\\
MRI-P-030501\\
ANL-HEP-PR-03-036\\
\end{flushright}

\vskip 3.5cm

\begin{center}
{\Large \bf Hybrid Inflation and Brane-Antibrane System}\\

\vspace*{6.0ex}

{\large \rm Debajyoti Choudhury$^{a,b}$, Debashis Ghoshal$^{a}$,\\
\vspace*{1.0ex}
Dileep P.\ Jatkar$^{a}$ and Sudhakar Panda$^{a,}$\footnote{{\tt
debchou, ghoshal, dileep, panda@mri.ernet.in}}}

\vspace*{3.5ex}

$^{a}$ {\large \it Harish-Chandra Research Institute, Chhatnag Road, Jhusi,\\
                   Allahabad 211019, India}\\[2ex]

$^{b}$ {\large \it HEP Division, Argonne National Laboratory, 
                   9700 Cass Ave.,\\ 
                   Argonne, IL 60439, USA}

\vspace*{10.0ex}

{\bf Abstract}

\begin{quote}
We study a string theory inspired model for hybrid inflation in the
context of a brane-antibrane system partially compactified on a compact
submanifold of (a caricature of) a Calabi-Yau manifold. The interbrane
distance acts as the inflaton, whereas the end of the inflationary epoch 
is brought about by the rapid rolling of the tachyon. The number of 
e-foldings is sufficiently large and is controlled by the initial 
conditions. The slow roll parameters, however, are essentially determined 
by the geometry and have little parametric dependence. Primordial 
density fluctuations can be made consistent with current data at 
the cost of reducing the string scale.
\end{quote}
\end{center}

\newpage

\setcounter{page}{1}

\tableofcontents


\sectiono{Introduction}
The knowledge of the history of the observable Universe before the
epoch of nucleosynthesis is rather limited. However, it is widely believed
that an early era of cosmological inflation existed \cite{linkt}. 
The dynamics of the Universe was dominated by a homogeneous scalar field, 
called the inflaton field, during this era. The potential for the inflaton
field accounted for almost all of the energy density of the Universe.
The latter
decreased with time as the scalar field rolled slowly down the
slope of the potential. Apart from explaining the observed large scale 
homogeneity and isotropy of the Universe, the inflationary paradigm 
explains structure formation as having been seeded by tiny primordial
density fluctuations. String theory, being a consistent theory of quantum 
gravity, ought to play a crucial role in providing a viable theory of 
inflation. An interesting direction that has been followed during
the last couple of years is the study of inflation in the context of
annihilation of extended objects called Dirichlet branes. There are
unstable configurations of these branes in which various scalar fields arise
in the low energy effective actions. Among these are the
transverse scalars governing the motion of the branes, the `tachyon'
field corresponding to the instability of the configuration and moduli
fields describing the background geometry. One or more of these could play
important roles in inflation. An incomplete list of references
is \cite{Herdeiro,BMNQRZ,DvSo,BMQRZ,list,Tinflat,list1,KofLin,list2,CFM,list3,KKKK};
see also \cite{quevedo} for a recent review.

In this paper we study the annihilation of a brane-antibrane system
as a model of hybrid inflation\cite{LinHyb,linkt} inspired by string theory.
The basic idea was first proposed
in Ref.\cite{BMNQRZ} (see also \cite{DvSo}). In these models,
interbrane separation acts as the inflaton field, and the one
that causes exit from the inflationary universe is the
tachyon field. The latter is a scalar field corresponding to
the lowest level excitation of the open string which connects
the brane and the antibrane. It becomes tachyonic when
the interbrane separation reaches the string scale. We consider
a pair of Dirichlet six and anti-sixbrane in (a caricature of) 
a Calabi-Yau space. Three of the worldvolume directions of the 
branes are wrapped on a three dimensional torus $T^3$. The branes 
fill the spacetime we live in, but are separated in an internal
three space, which we take to be a three dimensional sphere
$S^3$ or a real projective three-space $RP^3$.
The inflaton is the four dimensional scalar field that
corresponds to the separation between the branes. There is
an attractive potential between the brane and the anti-brane,
which, at large distances, is given by the solution of the
Poisson equation on the transverse space $S^3$ (or
$RP^3$). This potential drives inflation.
However, when the branes come close enough, an instability
develops and the rolling of this tachyonic scalar field
causes the universe to exit from the inflationary phase.
We find that it is possible to obtain enough expansion,
although the density perturbations tend to be slightly higher, 
with reasonable input parameters (string coupling, string length 
and sizes of the internal space). We emphasize that ours is
{\em not} a `brane world model'. In our model, the universe, at 
the end of inflation and the subsequent tachyon condensation, 
settles to a closed string vacuum determined by the Calabi-Yau 
manifold (possibly with fluxes\cite{FluxComp}, so that we have 
minimal supersymmetry).

It has been suggested that this annihilation scenario could be 
modified by starting with different numbers of branes and antibranes, 
so that one finally ends up with one (or more) brane(s) on which we 
live\cite{BMNQRZ}. However, this would require one to start
with a net brane charge on the compact transverse space. This
is disallowed by the requirement charge neutrality on a 
compact space. 

The plan of the paper is as follows.
In the next section, we review some aspects of brane-antibrane
geometry and dynamics\cite{ASen}. In Sec.~3, we couple this system
to gravity and derive the equations of motion. These are then
analyzed numerically in Sec.~4.  In the final section, we
end with some comments.


\sectiono{Dynamics of brane-antibrane \&\ spacetime geometry}
Type II string theory has Dirichlet branes of all dimensions.
More specifically, a stack of parallel D$p$-branes for even
(respectively odd) values of $p$ are supersymmetric in type
IIA (IIB) in flat ten dimensional spacetime. These D-branes
satisfy BPS condition and break half of the 
supersymmetries\footnote{Type IIA string has non-supersymmetric, 
{\em i.e.}, non-BPS branes for all odd
values of $p$, and conversely in IIB.}. An antibrane is oriented
opposite to a brane, and breaks the other half. Together, a 
parallel brane-antibrane pair (or
a stack of them) give rise to a non-supersymmetric configuration.
There is an attractive force between branes and antibranes.
This is due to the exchange of massless graviton, dilaton and
Ramond-Ramond (RR) forms\footnote{For BPS, {\em i.e.,}
supersymmetric brane configurations, the attractive force due 
to graviton and dilaton is cancelled by the repulsive force due 
to RR gauge fields.} (that are responsible
for the charge of the branes), as well as the infinite tower of
stringy fields; all of which are closed string modes.

Let us now focus our attention on a brane-antibrane pair,
which we will simply
refer to as branes in the following. When the distance $r$ between
the branes is large (in units of the string length scale
$\sqrt{2\pi\alpha'}$), this force can be calculated
from an effective field theory, and is of Coulomb type
$\sim r^{p-8}$, where $r$ is the radial coordinate transverse
to the branes. This results in a potential for the open string
scalar fields $Y^i$, ($i=1,\cdots,9-p$, $r=2\pi\alpha' |Y|$)
on the $(p+1)$-dimensional worldvolume of the branes.

Below a critical value $\pi\sqrt{2\alpha'}$ of interbrane
separation, an instability develops. This manifests itself by
making the lowest scalar excitation $T$ on the open string (connecting
the branes) tachyonic. (At the critical separation, the energy
due to the stretching of the string is just enough to balance
the negative zero point energy.) This instability is captured
by a potential $V(T,Y)$. Since both $T$ and $Y^i$ are
excitations of the open string, their interaction is at the
tree level (classical) and dominates over the potential term
involving the $Y$s. This is because the latter arises from
exchange of gravitons etc.\ in the loops.
Moreover, at separations comparable to the string scale, all
the massive modes of the string need to be taken into account,
and the simple analysis based on the effective field theory
breaks down.

The spacetime geometry we are interested in is not the ten
dimensional flat space, but a product of (3+1)-dimensional
flat spacetime with a six dimensional internal space ${\cal M}$.
In the absence of any brane, type II theory has ${\cal N}=2$
supersymmetry in four dimensions when ${\cal M}$ is a manifold
of SU(3) holonomy, a Calabi-Yau space. (In \cite{BMNQRZ} the
internal manifold is taken to be a torus, which preserves all
the supersymmetries. Subsequently, Ref.\cite{BMQRZ} considered
an orbifold of torus which is a singular limit of a Calabi-Yau
space. However, in the latter, interbrane separation is not the
inflaton.)

There are a large number of moduli fields in any of these
compactification schemes.
However, recent developments\cite{FluxComp} show that almost
all these moduli can be frozen by turning on suitable
fluxes in ${\cal M}$, and/or taking an `orientifold
quotient'. What is more, the resulting theory has minimal,
{\em i.e.}, ${\cal N}=1$ supersymmetry. Be that as it may,
a detailed model is beyond the scope of the present work. We
work with the assumption that some mechanism is at place to
freeze the unwanted fields, so that the only closed string
fields are the graviton and the RR $(p+1)$-form.

We will further assume, following the picture advocated in
Ref.\cite{SYZ}, that the Calabi-Yau space is a three-torus
$T^3$ fibration over a three dimensional base space ${\cal B}$.
Among possible base spaces are those that could topologically
be the three-sphere $S^3$, or the projective space $RP^3$,
especially if an orientifold quotient is used. In other words, 
as far as the topology of ${\cal M}$ is concerned, it can locally 
be written as a product ${\cal M}=T^3\times S^3$ (though at 
special points of ${\cal B}$, the fibre $T^3$ may degenerate). 
A single D6-brane which fills our spacetime, preserves half of 
the supersymmetries if the compact part of its worldvolume is along
$T^3$\cite{BBS}. An anti-brane, as usual, preserves the other half 
and together they break all the supersymmetries. This configuration
is a non-supersymmetric excitation over the supersymmetric vacuum
determined by ${\cal M}$. {}From now on we 
will work with this system of D6-$\bar{D6}$ branes. As far as the 
base ${\cal B}$ is concerned, these branes are zero dimensional and 
sit at specific points. A realistic calculation would require the 
knowledge of explicit Calabi-Yau metrics, which unfortunately 
are not available. We will, therefore, consider a crude approximation
of a Calabi-Yau manifold by treating ${\cal M}$ to be 
$T^3\times{\cal B}$. Moreover, we will put a flat metric on 
$T^3$ and the standard round metrics on $S^3$ and $RP^3$, and 
neglect the back reaction of the branes on the metric of these
spaces. 

The potential energy of this system of branes comes from
two sources. First, there is a contribution from the tension
of the 6-branes wrapping $T^3$. This is
\[
{\cal E}_0=2{\cal T}_6V_\parallel,
\]
where ${\cal T}_6$ is the 6-brane tension and $V_\parallel$
is the volume of $T^3$. Secondly, we have the energy due to
the Newtonian and Coulombic interactions. If we assume that
the branes are far apart on ${\cal B}$, this can be obtained
by solving the Poisson equation. The strength
of this interaction is determined by 
$G_{N({\cal B})}=(2\pi g_s^2)(2\pi\alpha')^4/8V_\parallel$, and the
interaction energy is
\begin{equation}\label{xion}
{\cal E}_{\hbox{\small int}} = 2{\cal T}_6V_\parallel\,
{g_s\sqrt{2\pi\alpha'}\over 8\sqrt{2\pi}}\, G(r) \equiv
2{\cal T}_6V_\parallel\,\Phi(r),
\end{equation}
where, $r$ is the geodesic distance between the branes and
\begin{equation}\label{green}
G(r) = \left\{
\begin{array}{ll}
\left[(r-\pi R_S)\cot\left(
{r\over R_S}\right)-R_S\right]/4\pi^2R_S^2,\quad &\hbox{\rm for }S^3\\
{}&{}\\
(2r-\pi R_P)\cot\left({r\over R_P}\right)/4\pi^2R_P^2,
\quad &\hbox{\rm for }RP^3,
\end{array}
\right.
\end{equation}
is the solution of Poisson equation obtained by following
the general procedure outlined in \cite{Helgason}. As mentioned
earlier, we have assumed the standard metric and ignored 
back-reaction due to the branes. 

The effective field theory in (3+1)-dimensional spacetime has
three scalar fields corresponding to the transverse motion on
${\cal B}$. The second contribution to the potential energy
described above gives rise to a potential term
$2{\cal T}_6V_\parallel\Phi(Y)$ for
the field $Y$ which corresponds to the interbrane separation
in ${\cal B}$. To get $\Phi(Y)$, one substitutes
$r=2\pi\alpha'Y$ in eqns.\refb{xion} and \refb{green}.
Notice that, while the interaction
term  ${\cal E}_{\hbox{\small int}}$ is due to (6+1)-dimensional
gravity,  the size of the internal space being small, its effect
is to provide a potential term for the field $Y$ in (3+1)
dimensions.

Incorporating the kinetic energy of $Y$, which we recall is of
the Born-Infeld form\cite{BI}, we can write the
lagrangian for the field $Y$ and the `tachyon' field $T$ as
\begin{equation}\label{matteraction}
{\cal L}_{\hbox{\small matter}} =
- 2T_6V_\parallel\;\left(V(T,Y)\sqrt{1-(2\pi\alpha')^2(
2f_c\partial_\mu T\partial^\mu\bar T +
\partial_\mu Y\partial^\mu Y)} + \Phi(Y)\right),
\end{equation}
where, $f_c=2\ln 2/\pi$\cite{Tinflat},
and $V(T,Y)$ is a potential for the tachyon field alluded to
earlier. An expression for this potential can be proposed
by a straightforward generalization of the result derived
in boundary string field theory (BSFT)\cite{BSFT}
\begin{equation}\label{bsftpot}
V(T,Y) = \exp\left(2\pi\alpha'\left(Y^2-{1\over 2\alpha'}
\right)|T|^2\right).
\end{equation}
This form of the potential can be motivated in the following way. When
the interbrane separation is zero, this potential matches with that
proposed by boundary string field theory. For small values of tachyon,
by expanding the exponential it is easy to see that it correctly
produces the tachyon mass formula.

The potential \refb{bsftpot}, however, does not give the expected
asymptotic behaviour for the tachyon\cite{senroll,senFT}. A better choice
follows from the recent proposal of Ref.\cite{KKKK,LaLiMa}
\begin{equation}\label{maldapot}
V(T,Y) = \sec\left(2\pi\alpha'|T|
\sqrt{Y^2-{1\over 2\alpha'}}\right).
\end{equation}
Both the functions have the property that for $Y^2> 1/2\alpha'$,
the mass-square of the field $T$ is positive, indeed the potential
is extremely steep for large separation of branes. On the other
hand, it becomes tachyonic below the critical separation.
We find that our conclusions are largely independent of
the two possible choices.


\sectiono{Coupling to gravity}
We will now couple the above field theory of scalars $T$ and $Y$
to (3+1)-dimensional gravity. We will assume a
Friedmann-Robertson-Walker metric,
\[
ds^2=-dt^2+a^2(t)\left(dx^2+dy^2+dz^2\right),
\]
(with zero spatial curvature for simplicity). Since we are
primarily interested in the time evolution of the fields, let us
further assume $T=T(t)$ and $Y=Y(t)$, {\it i.e.}, both the tachyon $T$
and the scalar $Y$ are functions only of time.

The lagrangian that describes the dynamics of this universe is
\begin{equation}\label{lagrangian}
{\cal L} = {3\over \kappa^2}(a{\dot a}^2 + a^2\ddot a) -
2{\cal T}_6V_\parallel\, a^3
\left(V(T,Y)\sqrt{1-(2\pi\alpha')^2(2f_c|\dot T|^2 +
{\dot Y}^2)} + \Phi(Y)\right) \ .
\end{equation}
The four dimensional gravitational constant $\kappa^2$ is obtained in 
terms of the ten dimensional coupling constant $\kappa_{(10)}^2$ through 
\begin{equation}\label{kappasq}
{1\over\kappa^2} = {V_\parallel V_\perp\over\kappa_{(10)}^2}
= 
               {2\pi^2 \hat R_T^3\hat R_P^3\over g_s^2(2\pi\alpha')}
               = 2{\cal T}_6V_\parallel\,{\pi^{3/2}\hat R_P^3\over
                       \sqrt{2} g_s}\,(2\pi\alpha')   
                           \qquad  {\rm for }\;  RP^3, 
\end{equation}
where, $\hat R_P = R_P/\sqrt{2\pi\alpha'}$ is the radius measured in
string units. (The value of $\kappa^2$ for $S^3$ base space is twice 
that of $RP^3$.) The ratio of the string scale to the four
dimensional Planck scale that follows from the above is
$(m_s/m_P)^2 = g_s^2/8\pi^2\hat R_T^3\hat R_P^3$.

The equations of motion are easily derived:
\begin{eqnarray}
{2\over3}\dot H + H^2
&=& {g_s\sqrt{2}\over 3 \, \pi^{3/2}\hat R_P^3(2\pi\alpha')}\left(
V\sqrt{{\cal B}_I} + \Phi\right),\qquad \mbox{\rm for } RP^3,
\label{aeom}\\
(2\pi\alpha')^2\ddot Y &=& -\,{\cal B}_I\left({1\over V}{dV\over dY}
+ (2\pi\alpha')^2 3H\dot Y + {1-(2\pi\alpha')^2{\dot Y}^2\over
\sqrt{{\cal B}_I} V}{d\Phi\over dY}\right),\label{yeom}\\
2f_c\ddot T &=& -\,{\cal B}_I\left({1\over (2\pi\alpha')^2\,V}
{dV\over dT} + 6Hf_c\dot T - {2f_c\dot T\dot Y\over\sqrt{{\cal B}_I} V}
{d\Phi\over dY}\right),\label{teom}
\end{eqnarray}
where, $H=\dot a/a$ and ${\cal B}_I = 1-(2\pi\alpha')^2
(2f_c|\dot T|^2 + {\dot Y}^2)$. In addition, we have
\begin{equation}\label{hsquare}
H^2 = {1\over 3}\kappa^2\,\rho = {2\over 3}
\kappa^2{\cal T}_6V_\parallel\,\left(V/\sqrt{{\cal B}_I}+
\Phi\right).
\end{equation}
As a matter of fact, the above equations simplify in practice.
At early times, when the separation between the branes is large,
the steep potential $V$ (eqn.\refb{bsftpot} or \refb{maldapot})
keeps $T$ pinned at $T=0$, thereby yielding $V=1$.
Once the tachyon is excited, the potential $\Phi$ can be neglected.
(Recall that it is due to one loop effect, while the interaction
between tachyon and $Y$ is at tree level.) In our numerical
computation, we take care of this by freezing $\Phi$ at the
point where the field $T$ is excited.

The overall potential seen by the field $Y$ at the early
stages of evolution, and indeed until the excitation of the
scalar $T$, is
\begin{equation}\label{effYpot}
{\cal V}_{\hbox{\small eff}}(Y) = 2{\cal T}_6V_\parallel\,\left(1 +
\Phi(Y)\right) = 2{\cal T}_6V_\parallel\,\left(1 +
{g_s\sqrt{2\pi\alpha'}\over 8\sqrt{2\pi}}\,G(2\pi\alpha' Y)\right).
\end{equation}
For large brane separations comparable to the radius of ${\cal B}$,
this potential goes as ${\cal V}_{\hbox{\small eff}}
\sim 1 - A\left(Y/\hat R\right)^2$. Although this is not flat,
we will see that slow roll conditions are satisfied. 
Note that the behaviour of ${\cal V}_{\hbox{\small eff}}$ is
different from that of \cite{BMNQRZ}, where it is of the form
$a - bx^4$.


\sectiono{Time evolution and Y-driven inflation}
The equations of motion \refb{aeom}--\refb{hsquare} are
unfortunately not amenable to analytic solution. We therefore
proceed to solve them numerically. It has already been noticed
in a similar set up in Ref.\ \cite{BMNQRZ} that, in order to get
sufficient inflation, the branes should initially be as far as
apart as possible in the compact space. For
the transverse space $S^3$ or $RP^3$, this means that they must be close
to a pair of antipodal points to begin with. To be concrete, we will
present most of our numerical results for $RP^3$ and comment 
on the (mostly quantitative) differences for the $S^3$ at the very end.

To begin with, we also assume
that the branes start with zero initial velocity, so once they
start moving, they approach each other head on. These are the
initial conditions that we use for $Y$. The field $T$, at this point, 
is in an extremely steep potential, and therefore it is a very massive
one. Fluctuations of $T$ in this case would cost a lot of energy and
within our framework it makes sense to set $T$ to
zero\footnote{As we did for massive string states. In spite of
this, it may be instructive to look at the fluctuations of $T$ at
the initial epoch, and we will return to this point later in this
section.}.

As the branes approach each other, $T$ starts becoming lighter and
when the interbrane separation nears the critical value
$Y\sim 1/\sqrt{2\alpha'}$, the dynamics of $T$ becomes important.
Close to this transition point, we mimic the fluctuation
of $T$ by transferring a small part of the kinetic energy of
$Y$ to $T$ at
\begin{equation}
Y_c = c_Y/\sqrt{2\alpha'}, \qquad (1.0\lsim c_Y\lsim 1.5).
\end{equation}
The fluctuation sets off an oscillation of $T$.
Once the separation goes below
the critical value, $T$ becomes tachyonic and starts to roll
down the potential $V(T,Y)$. We have already seen that the
interaction between $T$ and $Y$ is at tree level. Therefore,
once $T$ is excited, it is a valid approximation to neglect the
potential $\Phi(Y)$. In our solution, we take care of this by
freezing the value of $\Phi$ to $\Phi(Y_c)$. Strictly speaking,
for $Y\sim Y_c$, we are in a stringy regime, and the
low energy effective field theory description really breaks down.
However, once the tachyon starts rolling, its dynamics dominates.
An effective field theory is once again a good description\cite{ASeff}.

\begin{figure}[!ht]
\begin{center}
\centerline{\hspace*{18ex} 
            \epsfxsize=3.5in \epsfysize=1.7in \epsfbox{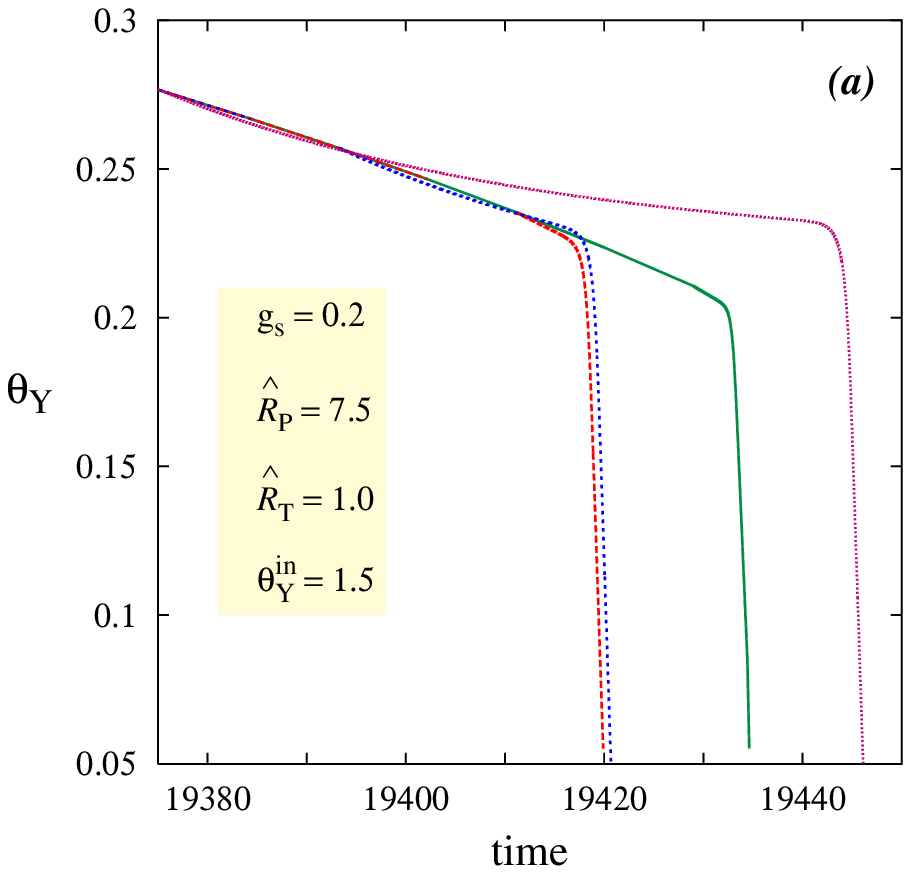} 
            \hspace*{-6ex} 
            \epsfxsize=3.5in \epsfysize=1.7in \epsfbox{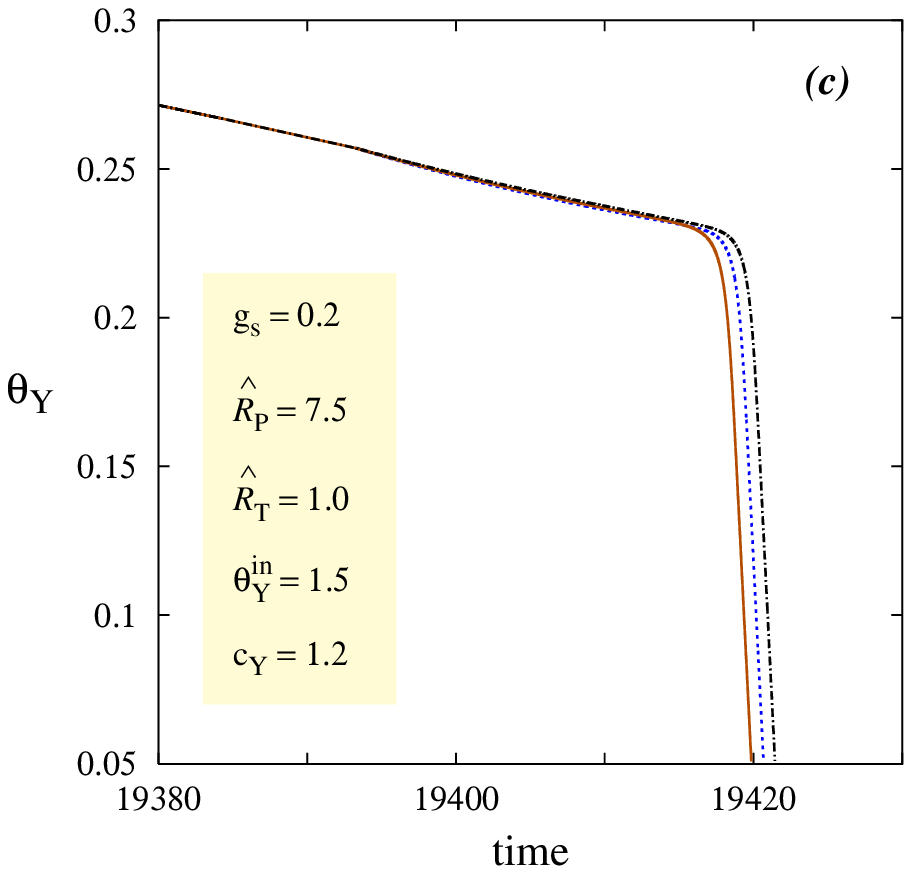}}
\vspace*{8ex}
\centerline{\hspace*{18ex}
            \epsfxsize=3.5in \epsfysize=1.7in \epsfbox{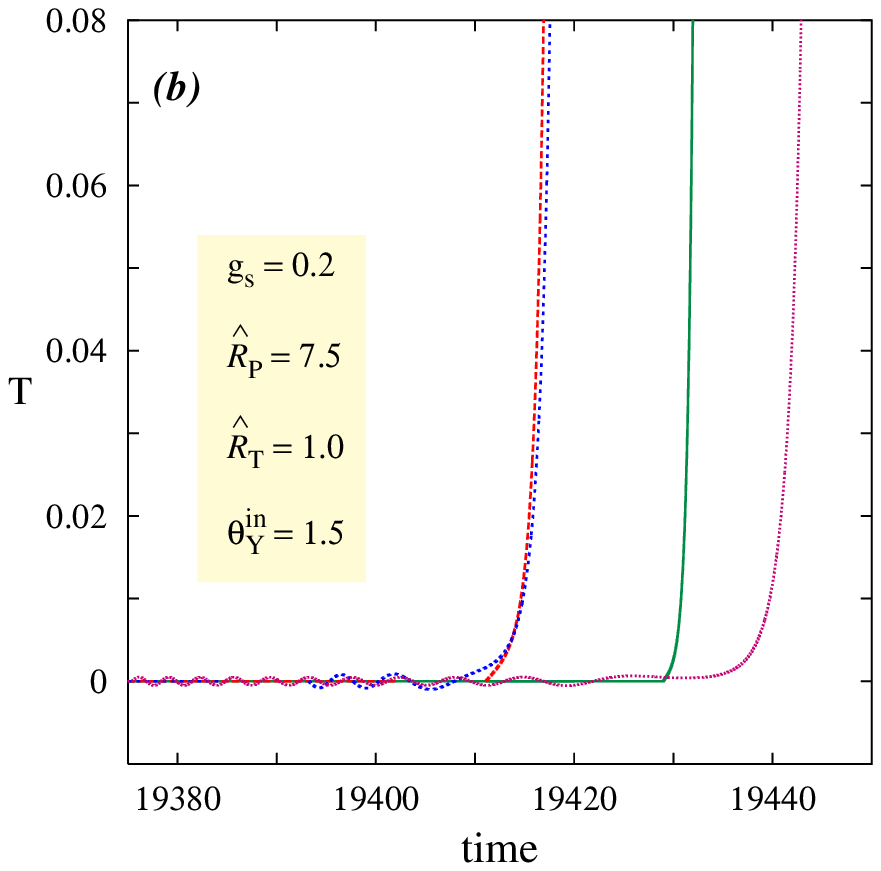} 
            \hspace*{-6ex} 
            \epsfxsize=3.5in \epsfysize=1.7in \epsfbox{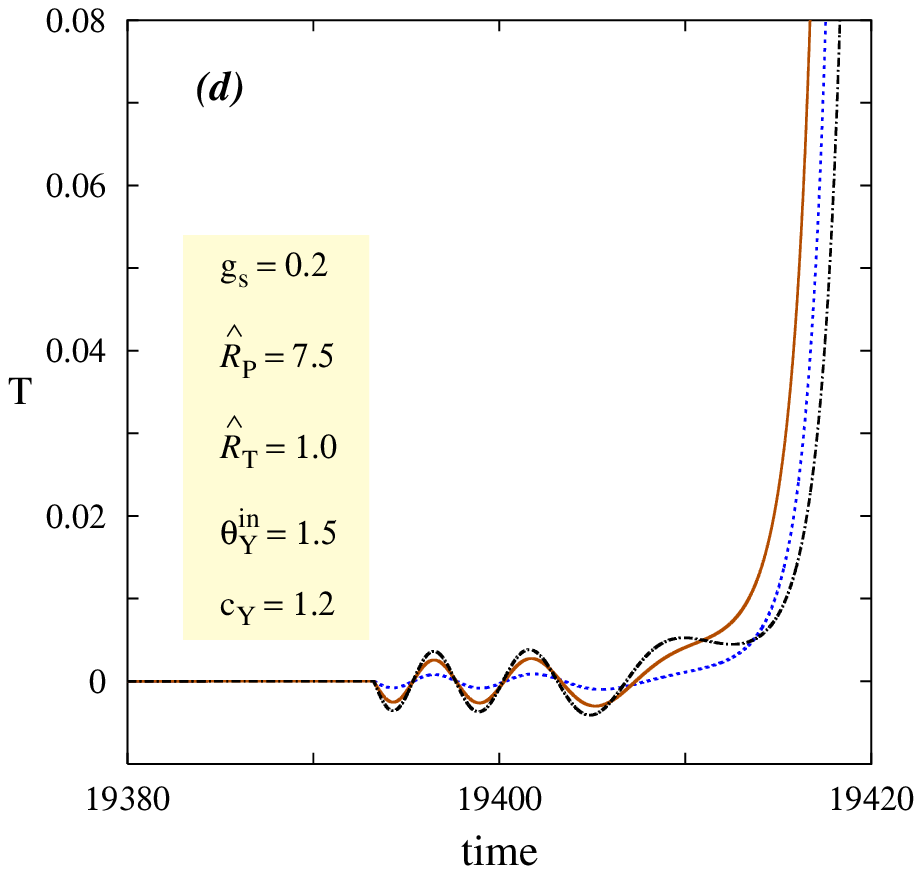}}
\vspace*{3ex}
\caption{\em The left panels depict the time evolution of the scalar fields
  $\theta_Y$ and $T$ subsequent to the transfer of kinetic energy
  from the former to the latter, for different transfer locations. The curves
  range over $c_Y = $ 1 to 1.5 (from right to left).
  In each case, 1\%\ of the energy is transferred.
  The panels on the right concentrate on a single location ($c_Y = 1.2$)
  but for different amounts of energy transfer (1\%\ to 20\%).
  }
\label{fig:kick}
\end{center}
\end{figure}

It might be argued that the parameter $c_Y$ introduces an unnecessary
element of arbitrariness, one that could, in principle, have been
eliminated had we engaged in a full string theoretic calculation. This
exercise being intractible, we find it instructive to examine, at this
stage, the dependence of the time evolution of the fields $T$ and $Y$  on
$c_Y$. For this purpose, and even for the rest of the numerical analysis,
it is convenient to consider the normalized field $\theta_Y$:
\begin{equation}
\theta_Y = {\sqrt{2\pi\alpha'}\,Y\over{\hat R}_P}
\equiv \frac{Y}{\hat R_P}, \qquad 0\le \theta_Y \le {\pi\over 2}.
\end{equation}
As an examination of Figs.~\ref{fig:kick}(a,b) shows, the subsequent
evolution does not suffer any qualitative change as $c_Y$ is varied
over a reasonably large range. Quantitatively, the onset of the rapid
roll down towards $\theta_Y=0$ is delayed or brought forward. This, 
in turn, determines the instant at which $T$ becomes tachyonic  and 
rolls down the potential. As we shall see below, most of the
inflation would have taken place before this transition point is reached.
Therefore, the number of e-foldings is largely independent\footnote{Of
course, if $c_Y$ were to be so large as to necessitate the freezing of
$\Phi(Y)$ relatively close to the antipodal points, the above arguments
clearly would not hold.} of $c_Y$.  In the remainder of this analysis
we shall use a representative value of $c_Y$, namely $c_Y=1.2$.

Another aspect that needs looking at is the dependence on the amount
of kinetic energy that is transferred from $Y$ to $T$. While the
heuristic picture presented above suggests that this fraction should
be a small one, in our investigations we have allowed for a fairly
large variation. The results are depicted in
Figs.~\ref{fig:kick}(c,d). Once again, it is obvious that no
qualitative change is brought about in the dynamics by transferring a
large amount of energy to the would-be tachyon. Hence, for the remainder
of this analysis we shall consider the case in which only 1\%\ of the
kinetic energy of $Y$ is transferred to $T$, at the point
$Y = c_Y/\sqrt{2 \alpha'}$.

\begin{figure}[!ht]
\leavevmode
\vspace*{-0.5truein}
\begin{center}
\epsfxsize=6.5truein
\epsfysize=3truein
\epsfbox{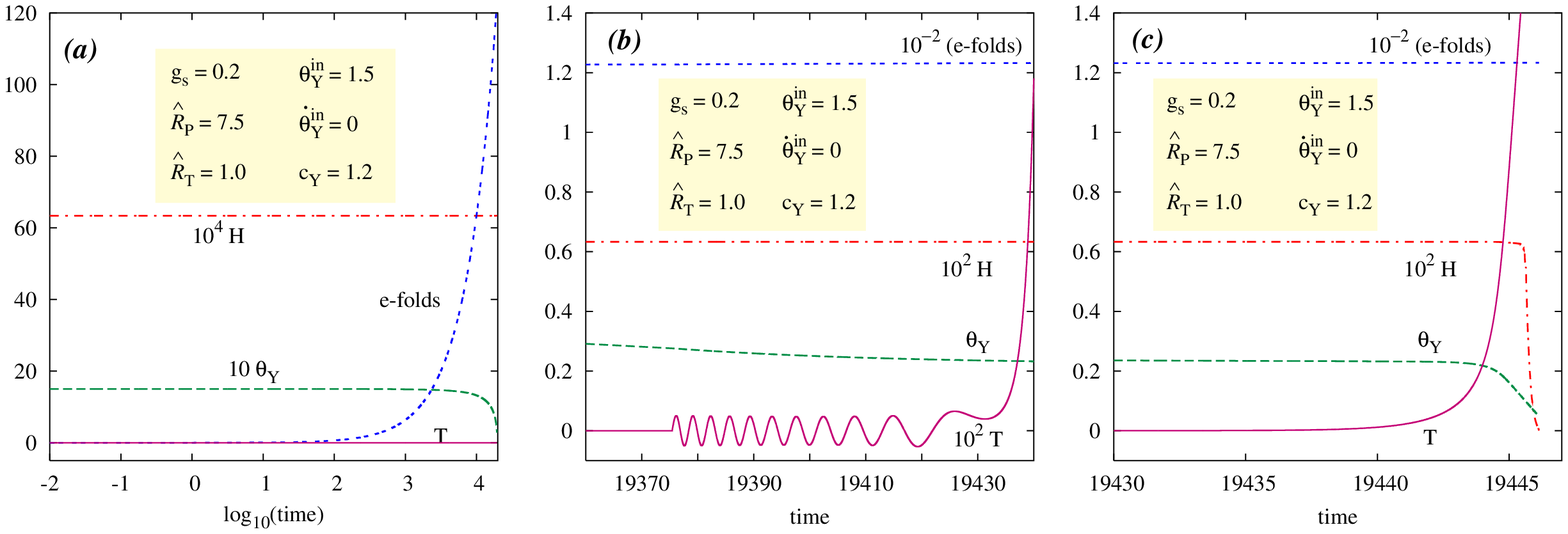}
\caption{\em The time variation of the fields $T$,
     $\theta_Y \: (\equiv Y / \hat R_P)$ and the Hubble parameter $H$.
     Also shown is the growth of the number of e-foldings
     with time. Panels {\em (b)} and {\em (c)} are zoomed-in views
     of tiny slices of {\em (a)}.
        }
    \label{fig:allinone}
\end{center}
\end{figure}

Let us now study, in detail, the complete time evolution of the system
for a representative set of parameter values. As has already been pointed
out, when the interbrane separation is large compared to the string length,
the field $T$ should not develop at all. In other words, at very early
epochs, one should start with $T_{\rm init} = 0,
\; \dot{T}_{\rm init} = 0$. The classical equations of motion \refb{teom}
immediately dictate that the field should continue to be firmly pinned at
$T = 0$. Consequently, as $Y$ rolls down the potential $\Phi(Y)$, it
exchanges energy only with gravity. With the effective potential
${\cal V}_{\hbox{\small eff}}(Y)$ being a slowly varying function of $Y$
(at least for $Y\sim R_{{\cal B}}$),
one expects that the initial evolution of $Y$ would be a very slow one.
With the time evolution of $H$ being governed by
eqn.\refb{hsquare}, this immediately translates to an even slower time
variation of $H$. Both these expectations are borne out by
Fig.~\ref{fig:allinone}(a). An immediate consequence is that the
universe experiences an (approximately) exponential growth rate with the
number of e-foldings, defined through
\[
   N_e (t) = \int_{t_0}^t dt \: H(t) \ ,
\]
growing almost linearly with $t$.

Once the field $T$ is excited, the dynamics undergoes a qualitative
change. The field $T$, still with a positive mass-square, can now
oscillate about its mean position (see Fig.~\ref{fig:allinone}(b)).
With $Y$ continuing its roll down towards $Y = 1/\sqrt{2\alpha'}$,
the effective mass of $T$ decreases with time, thereby increasing
the oscillation time period. As $Y$ falls below the critical value
$1/\sqrt{2\alpha'}$, $T$ becomes momentarily massless, then turns
tachyonic. This is reflected by a brief slowdown of the time evolution
of $T$, only to be followed by a rapid rolling down the potential.
It is interesting to note that even with an evolving $T$, the Hubble
parameter is almost  constant in time (at least, in the initial
phase). Only after $T$ has rolled down the potential sufficiently, 
does $H(t)$ start to suffer a (modest) decrease
(see Fig.~\ref{fig:allinone}(c)). 

\subsection{Conditions for slow roll}
Let us now discuss the conditions for slow roll. An investigation
of equations \refb{yeom} and \refb{hsquare} show that the slow
roll parameters are quite like the standard ones with
${\cal V}_{\hbox{\small eff}}$ in \refb{effYpot} as the potential.
In the regime in which we are interested in their values,
it is easy to check that the Born-Infeld form of the action does not
make much of a difference. Moreover, for most of the inflationary epoch,  
the evolution of $T$ can safely be neglected. The parameters that 
characterize slow roll, are
\begin{equation}
\varepsilon \equiv {1\over 4T_6V_\parallel\kappa^2}\,\left(
{{\cal V}'_{\hbox{\small eff}}\over
{\cal V}_{\hbox{\small eff}}}\right)^2,
\quad
\eta \equiv {1\over 2T_6V_\parallel\kappa^2}
\left({{\cal V}''_{\hbox{\small eff}}\over
{\cal V}_{\hbox{\small eff}}}\right),
\quad 
\xi \equiv \left({1\over 2T_6V_\parallel\kappa^2}\right)^2
\left({{\cal V}'_{\hbox{\small eff}}{\cal V}'''_{\hbox{\small eff}}\over
{\cal V}^2_{\hbox{\small eff}}}\right),
    \label{eps_eta}
\end{equation}
where, the factors of $2T_6V_{\parallel}$ can be attributed to
non-canonical normalization of the field $Y$.
Inflation is of slow roll type if $\varepsilon, |\eta|, |\xi| \ll 1$.
For the case of the $RP^3$ these expressions read
\begin{eqnarray}
\varepsilon 
     & = & \dis {g_s \over 8 (8\pi)^{3} \, (2 \pi)^{1/2} \, \hat R_P} \;
               \left[ \frac{ 2 \cot\theta_Y
                  - \left( 2 \theta_Y -\pi \right) \,
		\csc^2\theta_Y
		    }
                    { 1 + c_{RP} \, (2 \theta_Y -\pi) \,
		\cot\theta_Y }
		    \right]^2,\nonumber
\\[4ex]
\eta 
     & = & \dis {1 \over 32 \pi } \; 
               \frac{ \csc^2 \theta_Y \, 
		 \left( -2  + \left( 2 \theta_Y -\pi \right) \,
		   \, \cot\theta_Y \right)
		    }
                    { 1 +  c_{RP} \, (2 \theta_Y -\pi) \,
		\cot\theta_Y },\label{epseta_exp}
\\[4ex]
\xi 
     & = & \dis {\csc^2 \theta_Y \over 8 (16\pi)^{2}  } \;
          \frac{ 2 \cot\theta_Y
                  - \left( 2 \theta_Y -\pi \right) \,
		         \csc^2\theta_Y 
               }
	       { \left(1 + c_{RP} \, (2 \theta_Y -\pi) \,
		\cot\theta_Y \right)^2 
               } \; 
          \left[ 6\cot \theta_Y - \left( 2 \theta_Y -\pi \right) \,       
	     \left( 2  \cot^2 \theta_Y + \csc^2\theta_Y 
             \right)
         \right],\nonumber
\end{eqnarray}
where,
$
c_{RP} =  g_s/(8 (2\pi)^{5/ 2}\hat R_P),
$
and, for convenience, we have set $2\pi\alpha'=1$. It is worth 
pointing out that $\eta$ and $\xi$ are, for most part, 
independent of the parameters of the model\footnote{We thank
Nima Arkani-Hamed for emphasizing this point.}. Therefore, the slow
roll conditions are largely unaffected, leaving us the freedom to 
vary the parameters.

\begin{figure}[!ht]
\leavevmode
\vspace*{-0.2truein}
\begin{center}
\epsfxsize=6.5truein
\epsfbox{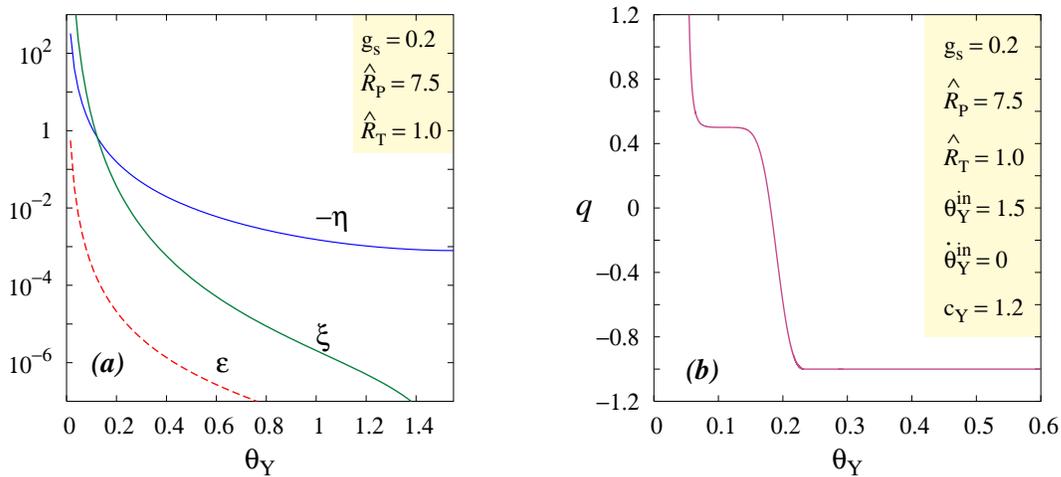}
\caption{\em {\em (a)} The slow roll parameters $\varepsilon$ and $\eta$
     as a function of the normalized field $\theta_Y$.
     {\em (b)} The deceleration parameter as a function of $\theta_Y$.
        }
    \label{fig:eps_eta}
\end{center}
\end{figure}

It is easy to
see that $\varepsilon\ll|\eta|$. As Fig.~\ref{fig:eps_eta}(a) shows,
$|\eta | \ll 1$ as long as $\theta_Y \gsim 0.5$. As can 
be checked, $\theta_Y \simeq 0.24$ corresponds to the point where
$T$ becomes tachyonic and the expansion of the universe starts to
slow down, {\em i.e.}, $H(t)$ starts decreasing with time. The 
spectral index 
\[
   n_s \equiv 1 - 6 \varepsilon + 2 \eta \ ,
\]
where the right hand side is to be evaluated at approximately 
60 e-folds before the end of inflation, 
is marginally below 1 and eminently consistent with the 
{\sc maxima, boomerang} and {\sc dasi} observations~\cite{spectral_tilt}. 
Note that, unlike in canonical hybrid inflation models, 
we have $\eta < 0$, a consequence of the shape of the potential 
$\Phi(Y)$. The running of the spectral index
\[
{d n_s \over d\,\ln k} = {2 \over 3} \left( \left( n_s - 1 \right)^2 - 
4 \eta^2 \right) + 2 \xi \sim   2\times 10^{-4},
\]
has a sign opposite to that preferred by the 
{\sc wmap} data\cite{wmap}. Note, however, that the extraction is 
beset with uncertainties, both observational and theoretical, and 
that our value is very well consistent at the $2 \sigma$ level.

Yet another relevant quantity is the deceleration parameter
$q$
\begin{equation}
   q = - \frac{\dot H}{H^2} - 1,
\end{equation}
which we have plotted in Fig.~\ref{fig:eps_eta}(b). That $q$ is almost
identically $-1$ for  $\theta_Y \gsim 0.24$ is but a restatement of the
exponential inflationary phase. On the other hand, $q>0$ signals
the end of the inflationary era and the beginning of the decelerating
phase. For the present choice of parameters, this occurs at
$\theta_Y \approx 0.2$, {\em i.e.}, a little after the tachyon
develops. In summary,  the slow roll expansion culminates in a tachyon 
driven fast roll one\cite{KofLin,Tinflat} and very quickly evolves 
into the decelerating phase. 

\subsection{Dependence on parameters \&\ Density perturbations }
\begin{figure}[!ht]
\leavevmode
\vspace*{-0.2truein}
\centerline{\hspace*{1ex} 
            \epsfxsize=3.5in \epsfysize=2.8in \epsfbox{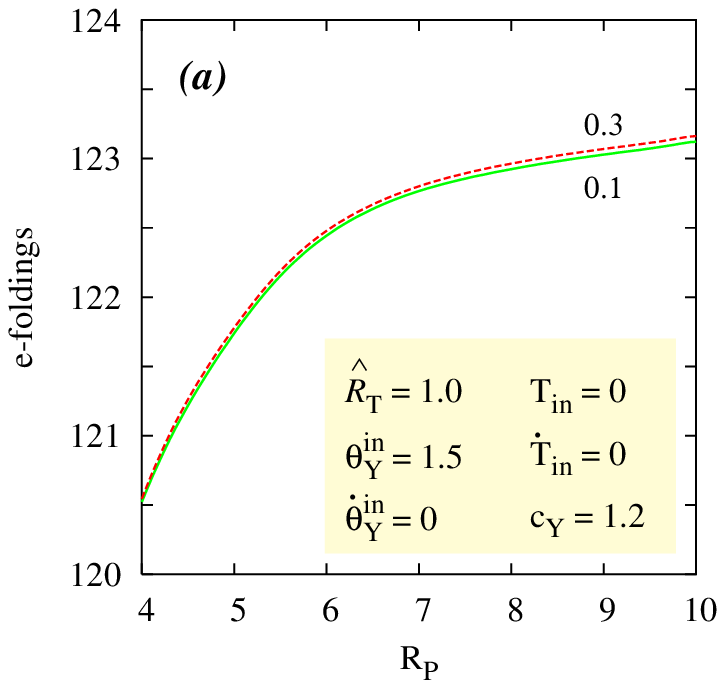} 
            \hspace*{-6ex} 
            \epsfxsize=3.5in \epsfysize=2.8in \epsfbox{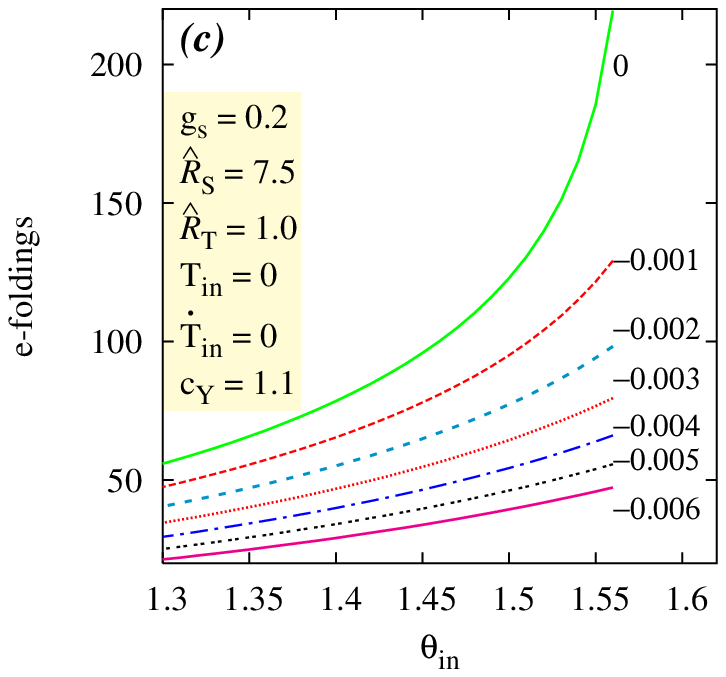}
           }
\vspace*{1ex}
\caption{\em {\em (a)} The dependence of the number of e-foldings on 
  $\hat R_P$, the radius of the projective three-space. The two curves 
  correspond to different values of the string coupling constant $g_s$.
   {\em (b)} The dependence of the number of e-foldings on the initial 
  position of the brane, on $RP^3$, with respect to the antibrane. 
  The different curves correspond to different values of the 
  initial inter-brane radial velocity.
        }
    \label{fig:param_dep}
\end{figure}

Until now, we have chosen to work only with a particular set
of model parameters. It is important that we check for the robustness
of our findings with variations in these. To start with, let us consider
the string coupling constant $g_s$ and the radius of the projective space
$\hat R_P$. As equations \refb{kappasq} and \refb{hsquare} imply, the 
Hubble parameter $H \propto \sqrt{g_s}$, and hence, all other 
things remaining the same, a smaller (respectively larger)
value for the latter would imply a smaller (larger) value for the 
number of e-foldings. There is, of course, 
an opposite effect on account of the change in the size of the
nontrivial term in the effective potential
${\cal V}_{\hbox{\small eff}}(Y)$---see eqn\refb{effYpot}--- as this
determines the rate at which $Y$ rolls down (and hence the 
length of the inflationary era). On convoluting the two effects, the 
resultant dependence is rather weak as is reflected
in Fig.~\ref{fig:param_dep}(a), where we display the variation in the
number of e-foldings with $\hat R_P$ for different values
of $g_s$. It should also be noted that the slow roll parameter
$\varepsilon \propto g_s$. However, since $\varepsilon$ is so very 
small during most of the inflationary phase (see Fig.~\ref{fig:eps_eta}), 
reasonable changes in $g_s$ would not result in any observable
effect.
The dependence on $\hat R_P$ is more subtle though. Since
$\kappa^2 \propto 1/{\hat R}_P^3$ (see eqn.\refb{kappasq}), it
would be tempting to conclude that $H(t) \propto 1/{\hat R}_P^{3/2}$
for most of the inflationary period and hence
$N_e \propto 1/{\hat R}_P^{3/2}$ as well. While the first
supposition is largely true, the second is not, as is clear from an 
inspection of Fig.~\ref{fig:param_dep}(a). In a large measure, this 
is due to the fact that the the distance to be traversed by the brane 
increases with $\hat R_P$. Furthermore, increasing $\hat R_P$ 
suppresses the slope of the potential ${\cal V}_{\hbox{\small eff}}(Y)$
resulting in a slower roll in the initial phase. In effect then, 
the amount of inflation is nearly independent of 
$\hat R_P$, though the duration of the inflationary phase is not. 

Is $\hat R_P$ then unmeasurable? Fortunately, observations guide us here. 
As the fields roll to their minima, adiabatic density perturbations
are caused. The amplitude of these perturbations can be 
estimated to be~\cite{linkt}
\beq
\barr{rcl}
\delta_H & \simeq & \dis
 {2T_6 V_\parallel\kappa^3\over 5\pi}\:
          { {\cal V}^{3/2}_{\hbox{\small eff}}(Y) \over 
                \left| {\cal V}'_{\hbox{\small eff}}(Y) \right| }
         \\[4ex]
& = & \dis { 2^{25/4} \sqrt{g_s}\over 5 \pi^{3/4} 
              {\hat R}^{5/2}_P }
                    \:
           { \left(1 + c_{RP} \, (2 \theta_Y -\pi) \,
		\cot\theta_Y \right)^{3/2}
              \over 
               \left| 
          2 \cot\theta_Y
                  - \left( 2 \theta_Y -\pi \right) \,
		\csc^2\theta_Y 
                   \right|}.
\earr
\eeq
It would immediately transpire that our result for $\delta_H$
(obtained with the default value for the radius $\hat R_P$) is 
three orders of magnitude larger 
than the {\sc cobe} result of $\delta_H \sim 2 \times 10^{-5}$. 
This, in a sense, was to be expected. With 
$\delta_H \propto \sqrt{\kappa^4 V /\varepsilon}$ for a generic theory, 
even a moderately small value of $\varepsilon$ would require that 
the appropriate value of the potential be much smaller than 
$1/\kappa^4$. With our theory being operative essentially at the 
Planck scale, this certainly is not true. 

On the other hand, the {\sc cobe} result could immdediately be used to 
fix $\hat R_P$. A better value of $\delta_H$ is obtained if $\hat R_P$ 
is scaled up by a factor of $\sim 16$. This has an immediate consequence 
in that the string scale comes out to be $\sim 6\times 10^{13}$ GeV
(assuming $\hat R_T =1$).

Increasing $\hat R_P$ has secondary effects. For one, the 
excitation of $T$ and, hence, the exit 
from inflation would now occur at even smaller values of $\theta_Y$. 
Since the slow roll parameters are essentially determined by 
$\theta_Y$, one might think that these grow significantly and 
thereby affect the spectral index inordinately. 
It can easily be checked though that this is not the case 
and that the differences (in either the value of $n_s$ or its 
running) would be barely noticeable at the scale of 
{\sc cobe} normalization. 

Let us now discuss the dependence on the initial separation and the
relative velocity of the branes on $RP^3$, which have been kept fixed
until now\footnote{The Ref.~\cite{BrGeWa} discusses the initial
conditions for brane inflation models.}. 
Placing them strictly at antipodal points with zero initial velocity 
would imply an equilibrium, though unstable, configuration. On the other 
hand, if the branes were to start far from the antipodal points, they would 
be feeling a relatively strong attractive force right from the beginning. 
Compounded with the fact that they would now have to traverse a smaller 
distance as well, it is easy to see that the number of e-foldings should 
reduce dramatically with the initial interbrane separation 
(see Fig.~\ref{fig:param_dep}(b)). For example, the 122 e-foldings
that we had obtained starting from $\theta_Y^{in} = 1.5$ increases 
to 220 if we start from $\theta_Y^{in} = 1.56$ and drops to only 
70 for $\theta_Y^{in} = 1.37$. Of course, if the brane had, in addition,
an initial radial velocity towards the antibrane, the reduction in 
the time of flight (and hence in $N_e$) is only helped 
(Fig.~\ref{fig:param_dep}(b)). Note, however, that with a 
nonzero initial velocity, the dependence of $N_e$ on $\theta_Y^{in}$ 
is weakened. This, again, is only to be expected. 

The only parameter that remains to be discussed is ${\hat R}_T$. As our 
equations of motion would testify, the dynamics 
is entirely independent of $\hat R_T$. The only place it enters is 
in the definition of the four dimensional gravitational coupling 
constant in terms of the ({\em a priori} unknown) ten dimensional one. 

\subsection{Effect of initial fluctuations of $T$}
\begin{figure}[!ht]
\leavevmode
\vspace*{-1.0truein}
\begin{center}
\epsfxsize=6.5truein
\epsfysize=4truein
\epsfbox{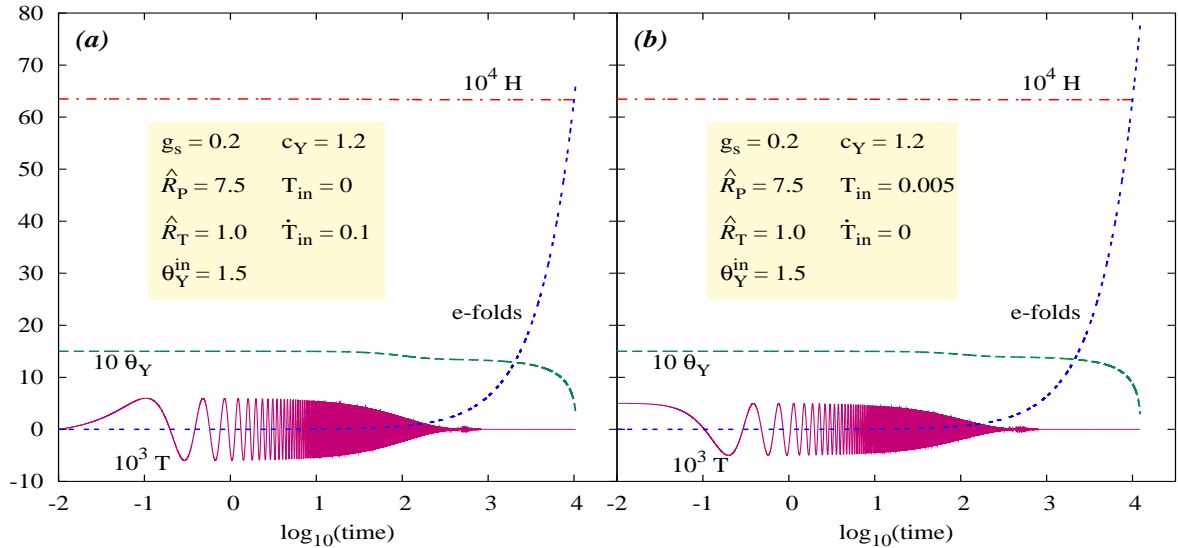}
\caption{\em The time variation of the the fields $T$, $\theta_Y$,
        Hubble parameter $H$ and the growth of the number of 
        e-foldings. The parameters are the same as in 
        Fig. \protect\ref{fig:allinone}, but initial fluctuations of $T$
        are included. {\em (a)} $T$ is given only a nonzero initial 
        velocity; {\em (b)} $T$ is given only a nonzero initial 
        displacement.
}  
\label{fig:initial_tach}
\end{center}
\end{figure}

Let us return to the issue of the fluctuations of the field $T$
at early stages of evolution. We may try to mimic random (quantum)
fluctuations by dispacing it from its minimum
or give it a non-zero velocity to start with. The consequent 
dynamics is portrayed in Fig.~\ref{fig:initial_tach}. While it may 
appear that the oscillation frequency is increasing rapidly with 
time, it is but an artifact of the choice of a logarithmic scale 
for time. In reality, the frequency {\em decreases} with time, as 
it should for a system losing energy. This is also reflected 
by the decreasing amplitude of the oscillation.
As the coupling of $T$ to $Y$ is stronger than that to the graviton, 
the energy that $T$ loses is essentially gained by $Y$, in the form
of kinetic energy. Consequently, $Y$ rolls faster, whereas $H(t)$ 
remains virtually unaltered.

\begin{figure}[!ht]
\leavevmode
\vspace*{-1.0truein}
\centerline{\hspace*{1ex} 
            \epsfxsize=3.5in \epsfysize=2.8in \epsfbox{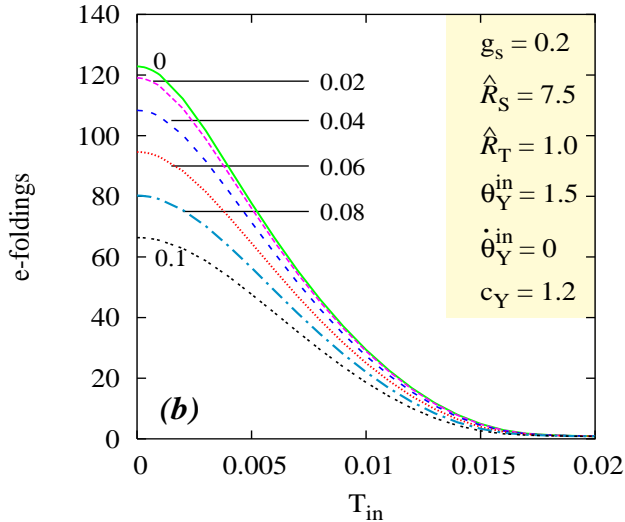} }
\vspace*{10ex}
\caption{\em The dependence of the number of e-foldings on the initial 
  value of the $T$ field. 
  The different curves correspond to different values of the 
  initial velocity $\dot T$.}
\label{fig:tach_dep}
\end{figure}

An immediate outcome of this faster rolling of $Y$ is the drop 
in the time allowed for inflation. With $H(t)$ remaining the same, 
this translates to a rapid drop in the number of 
e-foldings as is evinced by Fig.~\ref{fig:tach_dep}. 
However, as mentioned earlier, it is inconsistent to consider the 
fluctuations of the exteremely heavy field $T$ within the framework 
of a low energy effective theory. Moreover, if $T$ is excited, there
is no reason to exclude excitations of massive stringy modes, which
for large separation are lighter than the tachyon. While a proper 
resolution of this has to emerge from the full string theory, in 
a purely field theoretic hybrid inflationary model, this effect 
may reduce the number of e-foldings.

\sectiono{Discussion}

We have studied an hybrid inflationary scenario in which a
Dirichlet six- and antisix-brane, partially wrapped on a compact space,
annihilate. The geometry we have worked with is Minkowski spacetime times
a six dimensional compact space, which is a crude approximation to a
Calabi-Yau manifold. We also assume that some mechanism (like flux)
stabilizes unwanted closed string moduli and gives minimal supersymmetry.
Ours is not a brane world model in the sense that after the process
of brane annihilation, we are left with a purely closed string background
determined by the Calabi-Yau space (with fluxes). In particular, all
the standard model fields arise from the massless excitations of 
closed strings. 

\begin{figure}[!ht]
\leavevmode
\vspace*{-1.0truein}
\centerline{\hspace*{1ex} 
            \epsfxsize=3.5in \epsfysize=2.8in \epsfbox{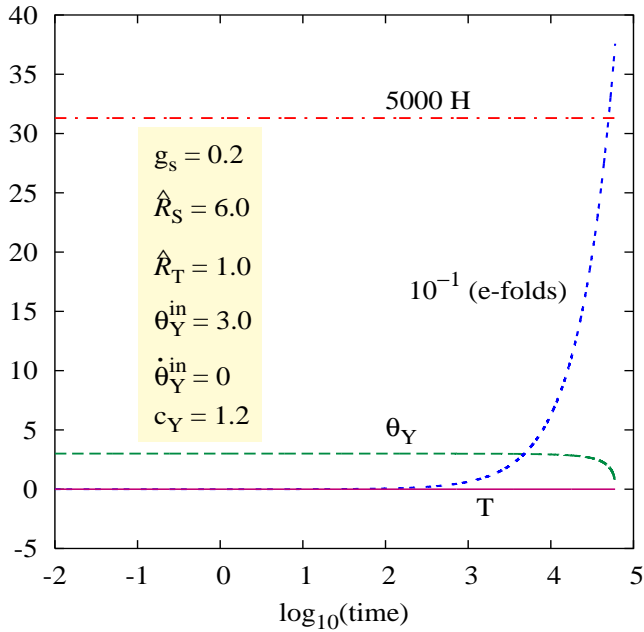} }
\vspace*{15ex}
\caption{\em The time variation of the fields $T$, 
     $\theta_Y \: (\equiv Y / \hat R_S)$ and the Hubble parameter $H$
     for the case of $S^3$. Also shown is the growth of the number of 
     e-foldings with time.
}
\label{fig:s3_infl}
\end{figure}

Most of the analysis in the paper is done for the case in which the 
space transverse to the branes is a projective three space $RP^3$. A
simpler choice would be the sphere $S^3$. As a matter of fact, there
is no qualitative difference between the two cases. Indeed most formulas
are almost identical except for some numerical factors, although the 
potentials \refb{green} that drive inflation differ in their details. 
It turns out that, for comparable choice of parameters, one gets much 
more (almost an order of magnitude increase in) e-foldings than $RP^3$. 
(Fig.~\ref{fig:s3_infl} shows the time evolution of some of the 
relevant quantities.) Notice that this gives one considerably more 
freedom in varying the initial conditions (interbrane separation and 
relative velocity), still keeping the amount of inflation within 
acceptable limits. 

\bigskip

One would like to know how the matter and radiation fields take
over at the end of inflation\footnote{The issue of reheating in
tachyonic inflation has been discussed in Ref.~\cite{CFM}. 
However, unlike ours, theirs is a brane world model.}. 
At first sight, it might seem that the runaway
potential of the tachyon does not allow for reheating to excite
the standard model fields. However, D-branes are excitations
of the closed superstrings, hence when they annihilate the
energy has to be converted into closed string modes. The details
of this process are just beginning to be 
addressed\cite{LaLiMa,Sbranes,SenOC}. However, the basic point
is that D-branes are classical sources of closed strings, and
D-branes with rolling tachyon are therefore time dependent 
classical sources. As in any field theory, coupling to time
dependent sources leads to particle production. In the case of
string theory, there are an infinite number of particles and
it turns out that, in a homogeneous process of brane decay,  heavy 
stringy modes of mass of order $1/g_s$ at zero spacetime
momenta are preferentially excited. This
is due to the fact that the tension of the D-brane is 
$\sim 1/g_s$. It is also consistent with the fact that these
heavy closed string modes have the same thermodynamic properties 
as the `tachyon matter'\cite{senroll,ASeff}. However, it is not yet
clear how and at what time scale these modes will decay into massless 
closed string states.

\bigskip

We have only considered a homogeneous decay of the complex 
tachyon. However, before the field $T$ turns tachyonic, its
potential flattens out allowing it to oscillate with large
amplitudes. Therefore, at different points of space, $T$ may
roll down along different directions in field space. Topological
defects may form in this process\cite{Feld}. It will be 
interesting to study this in detail. 

\bigskip

It is natural to wonder how the six-antisix brane pair come into
being and start in the way they do. Namely wrapped on supersymmetric
$T^3$-cycles of a compact Calabi-Yau manifold at nearly antipodal
points of a base manifold ${\cal B}$. Let us end by speculating
on a seemingly  natural way to obtain this within
type IIA string theory. Recall that this has space-filling D9-branes 
which are unstable. More specifically, there is a tachyonic scalar
on its worldvolume. It has been shown by Sen\cite{ASen} that all
the branes on type II string theory may be realized by appropriate
solitonic configuruation involving the tachyon (and gauge) fields 
of the unstable D9-brane. In particular, a six-brane is an 
't~Hooft-Polyakov monopole, (and an anti-sixbrane is an anti-monopole).
Let us 
start with an unstable D9-brane of type IIA theory in which the
geometry is four dimensional Minkowski times a compact Calabi-Yau
space (with fluxes). The tachyon condensation process leads to 
the formation of a D6-antiD6 pair as a monopole-antimonopole on 
the compact space ${\cal B}$. This appears to make sense for the
following reasons.
First, although the unstable D9-brane can decay into any of the lower
dimensional branes, the first ones that would be stable are the 
D6-branes. This is because a Calabi-Yau space does not have a 
non-trivial five-cycle on which the D8-branes can be wrapped to be 
stabilized and the D7-branes are
anyway unstable in type IIA theory. Secondly, on a compact space 
${\cal B}$, charge conservation would require that one always forms 
branes and antibranes in pairs. Therefore, as the local (radially
symmetric) formation 
of a brane tries to reduce energy, global issues (namely charge 
conservation) force us to end up with an unstable configuration 
again. Moreover, the formation of a monopole (sixbrane) at, say, the 
north pole of ${\cal B}=S^3$ will naturally produce an anti-monopole 
(anti-sixbrane) at the south pole. Of course, small irregularities 
would ensure that the branes do not form at {\em exact} antipodal points. 
{}From the four-dimensional point of view, the process outlined
above will also produce some inflation\cite{Tinflat} resulting in a 
couple of e-foldings. This {\em pre-inflation} is of fast roll
type\cite{KofLin,Tinflat} and might help in setting up appropriate
initial conditions\cite{LinFast}.

\bigskip

\noindent{\bf Acknowledgement:} It is a pleasure to thank 
Nima Arkani-Hamed, Gautam Mandal,
Shiraz Minwalla, Lubos Motl, Shinji Mukohyama and Ashoke Sen 
for valuable comments and discussions. 
DG and DPJ would like to thank the Harvard University,
especially Shiraz Minwalla, 
for hospitality during the final stages of this work. DC 
acknowledges the Department of Science and Technology, Government 
of India for financial assistance under Swarnajayanti Fellowship.



\newpage

\begin{thebibliography}{99}

\bibitem{linkt}
A. D. Linde, {\em Particle Physics and Inflationary Cosmology}, Harwood
Academic, Switzerland (1990);\\
E. W. Kolb and M. S. Turner, {\em The Early Universe}, Addision-Wesley (1990).
A.\ Liddle and D.\ Lyth, {\em Cosmological inflation and large scale
structure}, CUP (2000).

\bibitem{Herdeiro}
C.~Herdeiro, S.~Hirano and R.~Kallosh,
{\em String theory and hybrid inflation / acceleration}
JHEP {\bf 0112}, 027 (2001) [{\tt hep-th/0110271}];\\
K.~Dasgupta, C.~Herdeiro, S.~Hirano and R.~Kallosh,
{\em D3/D7 inflationary model and M-theory}, 
Phys.\ Rev.\ D {\bf 65}, 126002 (2002) [{\tt hep-th/0203019}].

\bibitem{BMNQRZ}
C.\ Burgess, M.\ Majumdar, D.\ Nolte, F.\ Quevedo, G.\ Rajesh
and R.\ Zhang, {\em The inflationary brane anti-brane universe},
JHEP {\bf 0107} (2001) 047 [{\tt hep-th/0105204}].  

\bibitem{DvSo}
G. Dvali, Q. Shafi and S. Solganik, {\em D-brane
inflation}, [{\tt hep-th/0105203}];\\
S. Alexander, {\em Inflation from D-anti-D brane annihilation}; 
Phys.\ Rev.\ {\bf D65} (2002) 023507 [{\tt hep-th/0105032}];\\
A.\ Mazumdar, S.\ Panda and A.\ Perez-Lorenzana,
{\em Assisted inflation via tachyon condensation}
Nucl.\ Phys.\ B {\bf 614} (2001) 101 [{\tt hep-ph/0107058}].

\bibitem{BMQRZ}
C. P. Burgess, P. Martineau, F. Quevedo, G. Rajesh and R. J. Zhang;
{\em Brane-Antibrane Inflation in Orbifold and Orientifold Models},
[{\tt hep-th/0111025}].

\bibitem{list}
G.~Gibbons, {\em Cosmological evolution of the rolling tachyon},
Phys.\ Lett. {\bf B537} (2002) 1, [{\tt hep-th/0204008}];
\\
M.~Fairbairn and M.~H.~Tytgat, {\em Inflation from a tachyon
fluid?}, Phys.\ Lett.\ B {\bf 546}, 1 (2002)
[{\tt hep-th/0204070}];
\\
S.~Mukohyama, {\em Brane cosmology driven by the rolling tachyon},
Phys.\ Rev.\ D {\bf 66}, 024009 (2002) [{\tt hep-th/0204084}];
\\
A.~Feinstein, {\em Power-law inflation from the rolling tachyon},
Phys.\ Rev.\ D {\bf 66}, 063511 (2002) [{\tt hep-th/0204140}];
\\
T.~Padmanabhan, {\em Accelerated expansion of the universe driven by
tachyonic matter}, Phys.\ Rev.\ D {\bf 66}, 021301 (2002)
[{\tt hep-th/0204150}];
\\
A.~V.~Frolov, L.~Kofman and A.~A.~Starobinsky, {\em Prospects and
problems of tachyon matter cosmology}, Phys.\ Lett.\ B {\bf 545},
8 (2002) [{\tt hep-th/0204187}].

\bibitem{Tinflat}
D.~Choudhury, D.~Ghoshal, D.~P.~Jatkar and S.~Panda, {\em On the
cosmological relevance of the tachyon}, Phys.\ Lett.\ B {\bf
544}, 231 (2002) [{\tt hep-th/0204204}].

\bibitem{list1}
G.~Shiu and I.~Wasserman, {\em Cosmological constraints on tachyon
matter}, Phys.\ Lett.\ B {\bf 541}, 6 (2002)
[{\tt hep-th/0205003}];
\\
H.~B.~Benaoum, {\em Accelerated universe from modified Chaplygin gas
and tachyonic fluid}, [{\tt hep-th/0205140}];
\\
M.~Sami, {\em Implementing power law inflation with rolling tachyon
on the brane}, [{\tt hep-th/0205146}];
\\
M.~Sami, P.~Chingangbam and T.~Qureshi, {\em Aspects of tachyonic
inflation with exponential potential}, Phys.\ Rev.\ D {\bf 66},
043530 (2002) [{\tt hep-th/0205179}].

\bibitem{KofLin}
L.~Kofman and A.~Linde, {\em Problems with tachyon inflation}, JHEP
{\bf 0207}, 004 (2002) [{\tt hep-th/0205121}].

\bibitem{list2}
T.~Mehen and B.~Wecht, {\em Gauge fields and scalars in rolling
tachyon backgrounds}, [{\tt hep-th/0206212}];
\\
K.~Ohta and T.~Yokono, {\em Gravitational approach to tachyon
matter}, Phys.\ Rev.\ D {\bf 66}, 125009 (2002)
[{\tt hep-th/0207004}];
\\
Y.~S.~Piao, R.~G.~Cai, X.~m.~Zhang and Y.~Z.~Zhang, {\em Assisted
tachyonic inflation}, Phys.\ Rev.\ D {\bf 66}, 121301 (2002)
[{\tt hep-ph/0207143}];
\\
G.~Shiu, S.~H.~Tye and I.~Wasserman, {\em Rolling tachyon in brane
world cosmology from superstring field theory},
[{\tt hep-th/0207119}];
\\
X.~z.~Li, D.~j.~Liu and J.~g.~Hao, {\em On the tachyon inflation},
[{\tt hep-th/0207146}];
\\
G.~N.~Felder, L.~Kofman and A.~Starobinsky, {\em Caustics in tachyon
matter and other Born-Infeld scalars}, JHEP {\bf 0209}, 026
(2002) [{\tt hep-th/0208019}];
\\
B.~Wang, E.~Abdalla and R.~K.~Su, {\em Dynamics and holographic
discreteness of tachyonic inflation}, [{\tt hep-th/0208023}];
\\
S.~Mukohyama, {\em Inhomogeneous tachyon decay, light-cone structure
and D-brane network  problem in tachyon cosmology}, Phys.\ Rev.\
D {\bf 66}, 123512 (2002) [{\tt hep-th/0208094}];
\\
M.~C.~Bento, O.~Bertolami and A.~A.~Sen, {\em Tachyonic inflation in
the braneworld scenario}, [{\tt hep-th/0208124}];
\\
J.-g.~Hao and X.-z.~Li, {\em Reconstructing the equation of state of
tachyon}, Phys.\ Rev.\ D {\bf 66}, 087301 (2002)
[{\tt hep-th/0209041}].

\bibitem{CFM}
J.~M.~Cline, H.~Firouzjahi and P.~Martineau, {\em Reheating from
tachyon condensation}, JHEP {\bf 0211}, 041 (2002)
[{\tt hep-th/0207156}].

\bibitem{list3}
C.-j.~Kim, H.~B.~Kim and Y.-b.~Kim, {\em Rolling tachyons in string
cosmology}, Phys.\ Lett.\ B {\bf 552}, 111 (2003)
[{\tt hep-th/0210101}];
\\
G.~Shiu, {\em Tachyon dynamics and brane cosmology},
[{\tt hep-th/0210313}];
\\
J.~S.~Bagla, H.~K.~Jassal and T.~Padmanabhan, {\em Cosmology with
tachyon field as dark energy}, [{\tt astro-ph/0212198}];
\\
Y.~S.~Piao, Q.~G.~Huang, X.-m.~Zhang and Y.~Z.~Zhang,
{\em Non-minimally coupled tachyon and inflation},
[{\tt hep-ph/0212219}];
\\
X.-z.~Li and X.-h.~Zhai, {\em The tachyon inflationary models with
exact mode functions}, [{\tt hep-ph/0301063}];
\\
G.~W.~Gibbons, {\em Thoughts on tachyon cosmology},
[{\tt hep-th/0301117}];
\\
M.~Sami, P.~Chingangbam and T.~Qureshi, {\em Cosmological aspects of
rolling tachyon}, [{\tt hep-th/0301140}];
\\
M.~Majumdar and A.~C.~Davis,
{\em Inflation from tachyon condensation, large N effects},
[{\tt hep-th/0304226}].

\bibitem{KKKK}
C.~Kim, H.~B.~Kim, Y.~B.~Kim and O-K.~Kwon, {\em Cosmology of rolling
tachyon}, [{\tt hep-th/0301142}].

\bibitem{quevedo}
F. Quevedo, {\em Lectures on String/Brane Cosmology}, [{\tt
hep-th/0210292}].

\bibitem{LinHyb}
A.~Linde, {\em Hybrid Inflation}, Phys.\ Rev.\ {\bf D49} (1993) 748
[{\tt astro-ph/9307002}].

\bibitem{FluxComp}
S.~Kachru, R.~Kallosh, A.~Linde and S.~P.~Trivedi,
{\em De Sitter vacua in string theory},
[{\tt hep-th/0301240}];\\
S.~Kachru, M.~B.~Schulz, P.~K.~Tripathy and S.~P.~Trivedi,
{\em New supersymmetric string compactifications},
JHEP {\bf 0303} (2003) 061
[{\tt hep-th/0211182}];\\
S.~Kachru, M.~B.~Schulz and S.~Trivedi,
{\em Moduli stabilization from fluxes in a simple IIB orientifold},
[{\tt hep-th/0201028}].

\bibitem{ASen}
A.\ Sen, {\em Stable nonBPS bound states of BPS D-branes}, 
JHEP {\bf 9808} (1998) 010 [{\tt hep-th/9805019}];\\
{\em Tachyon condensation on the brane anti-brane system}, 
JHEP {\bf 9808} (1998) 012 [{\tt hep-th/9805170}];\\
{\em Universality of the tachyon potential}, 
JHEP {\bf 9912} (1999) 027 [{\tt hep-th/9911116}].

\bibitem{SYZ} A. Strominger, S. T. Yau and E. Zaslow, {\em Mirror
Symmetry is T-Duality}, Nucl.\ Phys.\ {\bf B479} (1996) 243 [{\tt
hep-th/9606040}].

\bibitem{BBS} 
K.~Becker, M.~Becker and A.~Strominger,
{\em Five-branes, membranes and nonperturbative string theory},
Nucl.\ Phys.\ B {\bf 456} (1995) 130
[{\tt hep-th/9507158}].

\bibitem{Helgason}
S. Helgason, {\em Group and Geometric analysis}, AMS (2000).

\bibitem{BI}
A.\ Sen, {\em Supersymmetric world volume action for nonBPS D-branes}, 
JHEP {\bf 9910} (1999) 008 [{\tt hep-th/9909062}];\\
M.\ Garousi, {\em Tachyon couplings on nonBPS D-branes and
Dirac-Born-Infeld action}, Nucl.\ Phys.\ B {\bf 584} (2000) 284
[{\tt hep-th/0003122}];\\
E.\ Bergshoeff, M.\ de Roo, T.\ de Wit, E.\ Eyras and S.\ Panda,
{\em T duality and actions for nonBPS D-branes}, 
JHEP {\bf 0005} (2000) 009 [{\tt hep-th/0003221}];\\
J.\ Kluson, {\em Proposal for nonBPS D-brane action},
Phys.\ Rev.\ D {\bf 62} (2000) 126003 (2000)
[{\tt hep-th/0004106}];\\
A.~Sen, {\em Dirac-Born-Infeld action on the tachyon kink and vortex},
[{\tt hep-th/0303057}].

\bibitem{BSFT}
D.\ Kutasov, M.\ Mari\~ no and G.\ Moore, {\em Remarks on tachyon
condensation in superstring field theory}, [{\tt hep-th/0010108}];\\
P.\ Kraus and  F.\ Larsen, {\em Boundary string field theory of the D
anti-D system}, Phys.\ Rev.\ D {\bf 63} (2001)
106004 [{\tt hep-th/0012198}];\\
T.\ Takayanagi, S.\ Terashima, T.\ Uesugi, 
{\em Brane-anti-brane action from boundary string field theory},
JHEP 0103 (2001) 019 [{\tt hep-th/0012210}];\\
D.\ Ghoshal, {\em Normalization of the Boundary Superstring Field 
Theory Action}, {\tt hep-th/0106231}.

\bibitem{senroll}
A.\ Sen, {\em Rolling Tachyon}, JHEP {\bf 0204} (2002) 048 
[{\tt hep-th/0203211}].

\bibitem{senFT}
A.\ Sen, {\em Field theory of tachyon matter}, Mod.\ Phys.\ Lett.\ {\bf A17}
(2002) 1797 [{\tt hep-th/0204143}].

\bibitem{LaLiMa}
F. Leblond and A.W.Peet, {\em SD-brane gravity fields and rolling
tachyons},
[{\tt hep-th/0303035}];\\
N. Lambert, H. Liu and J. Maldacena; {\em Closed Strings from decaying
D-branes}, [{\tt hep-th/0303139}];\\
D. Kutasov and V. Niarchos; {\em Tachyon Effective Actions In Open String
Theory}, [{\tt hep-th/0304045}].

\bibitem{ASeff}
A.\ Sen, {\em Tachyon matter}, JHEP {\bf 0207} (2002) 065
[{\tt hep-th/0203265}].

\bibitem{spectral_tilt} A.T. Lee, et al., 
   {\em A High Spatial Resolution Anlysis of the MAXIMA-1 Cosmic 
          Microwave Background Data}, Astrophys.J. {\bf 561}, L1 (2001)
       [{\tt astro-ph/0104459}];\\
C.B. Netterfield et al., 
   {\em A measurement by BOOMERANG of multiple peaks in the angular 
          power spectrum of the cosmic mirowave background}, 
       Astrophys.J. {\bf 571}, 604 (2002) [{\tt astro-ph/0104460}];\\
C.Pryke, et al., {\em Cosmological Parameters Extraction 
from the First Season of Observations with DASI}, 
Astrophys.J. {\bf 568}, 46 (2002) [{\tt astro-ph/0104490}].

\bibitem{wmap}
C.~L.~Bennett {\it et al.},
  {\em First Year Wilkinson Microwave Anisotropy Probe (WMAP)
       Observations: Preliminary Maps and Basic Results},
[{\tt astro-ph/0302207}];\\
D.~N.~Spergel {\it et al.}, 
  {\em First Year Wilkinson Microwave Anisotropy Probe (WMAP)
    Observations: Determination of Cosmological Parameters}, 
[{\tt astro-ph/0302209}];\\
H.~V.~Peiris {\it et al.},
  {\em First Year Wilkinson Microwave Anisotropy Probe (WMAP)
         Observations: Implications for Inflation}, 
[{\tt astro-ph/0302225}].

\bibitem{BrGeWa}
R.~Brandenberger, G.~Geshnizjani and S.~Watson, {\em On the Initial
Conditions for Brane Inflation}, [{\tt hep-th/0302222}].

\bibitem{Sbranes}
A.~Maloney, A.~Strominger and X.~Yin, {\em S-brane thermodynamics},
[{\tt hep-th/0302146}];\\
D.~Gaiotto, N.~Itzhaki and L.~Rastelli, {\em Closed strings as 
imaginary D-branes}, [{\tt hep-th/0304192}].

\bibitem{SenOC}
A.~Sen, {\em Open and closed string from unstable D-branes}, 
[{\tt hep-th/0305011}].

\bibitem{Feld}
G.~Felder, J.~Garcia-Bellido, P.~Greene, L.~Kofman, A.~Linde and
I.~Tkachev, {\em Dynamics of symmetry breaking and tachyonic preheating},
Phys.\ Rev.\ Lett.\ {\bf 87} (2001) 11601;\\
L.~Pogosian, S.-H.~Tye, I.~Wasserman and M.~Wyman, {\em Observational
constraints on cosmic string production during brane inflation},
[{\tt hep-th/0304188}].

\bibitem{LinFast}
A.~Linde, {\em Fast-roll inflation},
JHEP {\bf 0111} (2001) 052 [{\tt hep-th/0110195}].

\end{thebibliography}
\end{document}